\documentclass[english,aps,floats,floatfix,footnotes,preprint]{revtex4} 
\usepackage[T1]{fontenc} 
\usepackage[latin9]{inputenc} 
\usepackage{amstext} 
\usepackage{graphicx} 
\usepackage{esint} 
\def\be{\begin{equation}} 
\def\ee{\end{equation}} 
 
\usepackage{babel} 
 
\begin{document}

\title{Spin constraints on nuclear energy density functionals}

\author{L.M.~Robledo} 
\email{luis.robledo@uam.es} 
\homepage{http://gamma.ft.uam.es/robledo} 
\affiliation{Departamento de F\'\i sica Te\'orica, M\'odulo 15, Universidad Aut\'onoma de 
Madrid, E-28049 Madrid, Spain}

\author{R.N. Bernard}
\affiliation{
Departamento de F\'\i sica At\'omica, Molecular y Nuclear, Universidad de Granada,
E-18071, Granada, Spain            
}

\author{G.F. Bertsch } 
\affiliation{Institute for Nuclear Theory and Dept. of Physics, Box 351560, University 
of Washington, Seattle, Washington 98915, USA} 
 
\begin{abstract} 
Abstract:  The Gallagher-Moszkowski rule in the spectroscopy of 
odd-odd nuclei imposes
a new spin constraint on the energy functionals for self-consistent
mean field theory.  The commonly used parameterization of the effective
three-body interaction in the Gogny and Skyrme families of energy 
functionals is ill-suited to satisfy the spin constraint.
In particular, the Gogny parameterization of the three-body interaction
has the opposite spin dependence to that required by the observed
spectra.  The two-body part has a correct sign, but in combination
the rule is violated as often as not.  We conclude that a new functional
form is needed for the effective three-body interaction that can take
into better account the different spin-isospin channels of the interaction.  
\end{abstract} 
\maketitle

Nuclear energy density functionals have reached a mature
state with respect to the parameterization of time-even fields required
for the Hartree-Fock-Bogoliubov theory of ground state properties of 
even-even nuclei.  A theory encompassing nuclei with odd numbers of 
nucleons requires a good understanding of the time-odd fields as well.
Some aspects of the time-odd fields have been examined in the literature
\cite{po85,pa76,do95,be02,po12},
but important experimental information has been ignored in determining
the parameters in the functionals.  In particular, the spin dependence
of the neutron-proton interaction is crucial to determine
ground-state spins of odd-odd nuclei.  We will show in this letter that
an energy functional from the Gogny family of functionals strongly 
violates an empirical rule for determine ground state spins.  The Gogny
functional has a very specific form for an effective three-body interaction
which automatically has the wrong sign for the spin dependence.  The 
other leading family of functionals, based on Skyrme's parameterization,
has the same form for spatial dependence of the effective 
three-body interaction, and is likely to have the similar difficulties.  Indeed, 
it was shown long ago that the contact parameterization could lead to
instabilities in nuclear Hartree-Fock theory\cite{pa76}.

The rule that should be respected was formulated by Gallagher and
Moszkowski (GM) \cite{ga58} for the quasiparticle angular momentum
couplings in strongly deformed odd-odd nuclei.  Under those conditions
the components of the 
angular momentum $K_p,K_n$ of the odd nucleons about the symmetry axis
are good 
quantum numbers.  The two possible relative spin orientations,
$K_p+K_n$ and $|K_p-K_n|$ give rise to two separate rotational bands
having band-head angular momentum $J = |K_p\pm K_n|$.
According to the rule \cite{ga58,bo76},
the oriention with parallel intrinsic spins is the lower energy
band.  
As documented in a review of the GM rule \cite{bo76},
there are only rare exceptions to the rule.

We have developed new code to find the Hartree-Fock-Bogoliubov minima
of the Gogny functional in axially symmetric nuclei, 
treating for the first time time-odd fields
including the spin-dependent ones
\cite{ro12}.  Applying the code to spin splittings in
deformed nuclei, we found that the predicted splittings
violated the GM rule more often than not.
In retrospect, the
result is not too surprising because as stated earlier none of the 
energy functionals in common use have been fitted to spin-dependent 
properties\footnote{ There has been some cognizance of the spin
properties when it was realized that some parameterizations lead to 
disasterous instabilities \cite{instability}.  
}.

We now examine the origin of the results.  It is useful to distinguish
the two-particle interaction and the three-particle interaction present
in the functionals. 
In principle there are enough
degrees of freedom in the parameterization of the two-particle 
interaction to take into account the GM splittings.  However, the
three-particle interaction is essential for nuclear saturation and, for
computational simplicity, it 
has a very constrained parameterization. Namely, 
it is a density-dependent contact interaction in both the 
Skyrme and Gogny functionals of the form
\be
t_3 (1 + x \hat P_\sigma) \delta(\vec r_1-\vec r_2) \rho((\vec r_1+\vec r_2)/2)^\alpha
\ee
in the standard notation\cite{RMP}.
It is further restricted to
the parallel-spin interaction ($x=-1$) in the Gogny functionals.  It must be
repulsive to saturate nuclear matter, but it can't have a significant
antiparallel-spin component because that channel requires
an attractive interaction overall to produce BCS pairing.

We now illustrate the problem with a well-known example, 
\def\lu{{$^{174}$Lu}}
the nucleus \lu.  The odd nucleons in ground band have angular momenta
and parities 
$(K_p,K_n) = (7/2^+,5/2^-)$ for the proton and neutron respectively.
These correspond to Nilsson orbitals $[404]\downarrow_p$ and
$[512]\uparrow_n$.
The spins are parallel for antiparallel orbital angular momentum, i.e.
$K = |K_p - K_n|$.  Indeed, the ground state band has $K^\pi = 1^-$
in agreement with the Gallagher-Moszkowski rule.  The other coupling
of angular momenta, $K = K_p+K_n = 6^-$, is associated with an excited
band with a band head at 171 keV excitation.  The experimental levels
are compared with the HFB calculations in Fig. 1.  We first show
the spectra of neighboring odd-A nuclei on the lefthand and 
middle panels.  In the middle one, the theory confirmed the ground band 
assignment of a quasiparticle in 
the $[5 1 2]\uparrow_n$ Nilsson orbital.  However, the theory does
not predict the correct ordering of the proton quasiparticle energies,
shown in the left-hand panel.  As a consequence, the $ [404]\downarrow_p
[512]\uparrow_n$ appear as excited states in the theoretical spectrum
of the \lu, shown in the right-hand panel.  One sees that the
level ordering is opposite to the experimental, with the $6^-$
band head below the $1^-$, thus violating the GM rule\footnote{The splittings has 
additional contributions besides the spin-dependent interactions.  However,
these are smaller \cite{bo76} and generally do not change the sign 
of the splitting.}.

To understand to theoretical splittings in more detail, we
separate three contributions:
1) the spin dependence of the two-body interaction, treating the
interaction in first-order perturbation theory
2) the spin dependence of the density-dependent interaction, again
treating it perturbatively
3) the many-body rearrangement effects associated with the wave function 
modifications in the two-quasiparticle state.  The three contributions
are $+188$ and $-291$ keV for the two-particle and three-particle
perturbative contributions, respectively.  The rearrange contribution
is $+44$ keV, giving a total splitting of $-61$ keV as shown in 
the level scheme in Fig. 1.  This should be compared with an empirical
value of $+114$, which is what is left of the observed splitting of 
$+171$ after the rotational effects have been removed \cite{bo76}.
Thus, as claimed earlier, the three-particle contribution has a bad
sign and here it even overwhelms the good sign of the two-particle
contribution.

\begin{figure}
\includegraphics [width = 11cm]{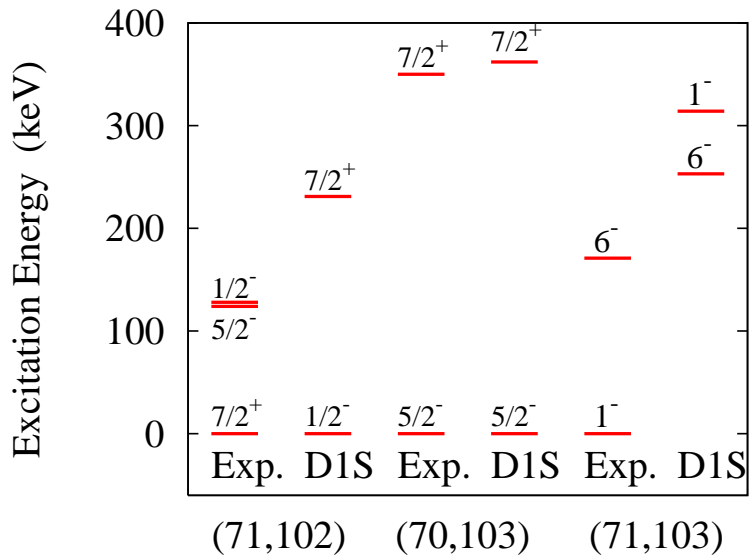}
\caption{\label{levels}  Low-lying band heads  in the spectra of the
nucleus \lu~and odd-A neighbors:  $^{173}$Lu (left); $^{173}$Yb (center)
\lu~ (right).  Due to the inversion of the lowest proton quasiparticle
energies, the ground-state doublet in
\lu~is not the lowest two-quasiparticle configuration in the
calculated spectrum.  Lower energy calculated configurations are
not shown.
}
\end{figure}

To see how general these findings are, we have performed this
analysis on 100-225 doublets in each of 15 nuclei in the deformed 
rare-earth region.  All of these nuclei have strong prolate
deformations.  The results are shown in Table I.
\begin{table}[htb]
\begin{center}
\begin{tabular}{|c|ccc|c|}
\hline
\hline
  & 2BP & 3BP   & FP & self-consistent \\
\hline
$^{164-168}$Ho  & 93\% & 8\% & 28\%& 45\% \\
$^{168-172}$Tm  & 97 & 4 & 26& 41 \\
$^{172-176}$Lu  & 97 & 4 & 28 & 40 \\
$^{180-184}$Ta  & 97 & 5 & 37 &30 \\
$^{184-188}$Lu  & 97 & 3 & 36 & 28 \\
\hline
\end{tabular}
\caption{\label{systematics}Theoretical spin splittings of neutron-proton configurations for odd-odd
nuclei in the rare earth region. For each nucleus, two-quasiparticle states
were constructed taking 10 to 15 proton quasiparticle orbitals  and a like
number of neutron quasiparticle orbitals.  The table shows the percentage
of the cases in which the calculated splitting agrees with the GM rule,
combining the results for several isotopes of each element.  
Columns labeled 2BP, 3BP, and FP show the perturbative results for 
the two-body interaction alone, 
the three-body interaction alone, and the full interaction treated 
perturbatively.  The last column shows the results of the fully
self-consistent calculation of the HFB minima.  The table shows the
results for the D1S interaction.  We also have calculated splittings
with the D1M interactions \cite{D1M} and found similar results.
}
\end{center}
\end{table}
These results confirm the statements made earlier that: the two-body
interaction has a correct sign; the three-body interaction has the
wrong sign; and the net sign with all the contributions is variable
and inconsistent with a general GM rule.

To gain a better understanding of the origin of the problem we briefly
review how the interaction energies are calculated using the one-body
densities of Hartree-Fock-Bogoliubov (HFB) theory.  
When time-reversal symmetry is broken the one body-density matrix can be decomposed 
as the sum of a time-even and a time-odd density. In the 
expression of the total energy there is a contribution 
which is quadratic in the time-odd term. Starting from an even-even HFB reference state, the 
blocking of a given quasiparticle leads to a non-zero time odd 
density matrix. The blocking of the time reversed state leads to the 
same time odd density but with opposite sign. To build the two 
configurations defining a GM pair a proton quasiparticle with 
quantum number $K_p$ and a neutron one with $K_n$ are blocked to 
obtain one of the states.  The other is obtained by blocking 
$(-K_p,K_n)$. In this way both the spin parallel and antiparallel 
configurations are considered\footnote{Which one is which depends on the 
decomposition of $K$ into orbital and spin components, $K=\ell_z + s_z$.}.
The time-odd proton densities of both configurations are the same in absolute 
value but have opposite signs whereas the two time odd neutron 
densities are the same. From these considerations it becomes clear 
that only those terms of the energy depending on the product of a 
time odd proton density times a time odd neutron one are 
contributing to the energy splitting of the doublet. Among the different
terms contributing to the energy in the Gogny interaction there are a 
few that do not contribute to the splitting, namely the Wigner term 
of the central potential, the Coulomb potential and the pairing channel
of the central potential. Among the remaining terms, the spin-orbit 
contribution is much smaller than the other two and will be omitted in the
discussion. Therefore the splitting
of the doublet is dominated by the central two-body and three-body
contributions.  We calculate the perturbative 
contribution to
the splitting starting a wave function at the HFB minimum of an even-even
nucleus.  The required quasiparticles are then blocked and the expection
value of the energy is calculated.  As an example, 
the three-body contribution to the splitting is given  by
\be
\Delta E = E(\uparrow,\uparrow) - E(\downarrow,\uparrow) 
\ee
$$
= 4 t_3 \int d^3 \vec{r} 
\rho^\alpha  \left(\rho^{p,odd}_{1/2, 1/2}  \rho^{n,odd}_{1/2, 1/2} 
+
\bar{\rho}^{p,odd}_{1/2, -1/2}  \bar{\rho}^{n,odd}_{1/2, -1/2}\right)
$$
where $\rho$ is the ordinary density, a function of $\vec r$ alone.
The needed time-odd component of the density matrix $\rho^{t,odd}_{s, s'} $ 
depends on nucleon type $t$ and spin projection $s,s'$ as well.
The bar denotes the modulus of a
(complex) density.
If the blocked quasiparticle is BCS-like (i.e. linear combinations of
creation and annihilation canonical basis states) then the time-odd 
density $\rho^{t,odd}_{s,s'}$
is diagonal in the canonical basis with zeroes in the diagonal except for
the blocked orbital quantum number where it takes the value $\pm 1/2$ depending
on whether the spin $\sigma$ of the blocked orbital points up or down.
In this very specific case only taking place at the first iteration (first order)
the density $\bar{\rho}^{t,odd}_{1/2,-1/2}$ is zero and $\rho^{t,odd}_{1/2,1/2}$ equals
$\sum_q|\varphi_{t,q} (\vec{r})|^{2}/2(-1)^{\sigma-1/2}$. Therefore $\Delta E$ as 
defined above is positive for parallel spins and negative for antiparallel
ones, just the opposite of the GM rule. 
In the actual HFB calculation the blocked     
quasiparticle may have a mixture of the two spin orientations
and the simple argument given above may fail.  This occurs for some
configurations treated in the Table.
 
It is also of interest to examine the various interactions in a momentum
space representation.  The relevant plane-wave matrix elements of the 
two-body Gogny
interaction are given by
\be
<q|V|q>_{\uparrow\uparrow}^{np} =\pi^{3/2}\sum_{i=1}^2 \mu_i^3
\left( W_i +B_i + (H_i+M_i)e^{-(q \mu_i)^2}
\right)
\ee
\be
<q|V|q>_{\uparrow\downarrow}^{np} =\pi^{3/2}\sum_{i=1}^2 \mu_i^3
\left( W_i  + M_ie^{-(q \mu_i)^2}
\right)
\ee
in the notation of Ref. \cite{Gogny}.  The momentum $q=|k_n-k_p|/2$ is the
relative momentum of the two nucleons. 

We now make a qualitative connection with the energy difference for the two spin
couplings of nucleons, changing the momentum of the proton at the 
same time as its spin is flipped.  The difference is\footnote{There is
another spin-isospin combination of the interaction term corresponding
to the Landau parameters for the interaction\cite[App. E]{be02}.  It is quite different
from Eq. (\ref{DeltaV}) .}
\be
\label{DeltaV}
\Delta v^{2b} = 
\pi^{3/2}\sum_{i=1}^2 \mu_i^3 ( B_i + (H_i+M_i) 
e^{-(q_p \mu_i)^2}
-M_i  e^{-(q_a \mu_i)^2} )
\ee
with $q_p,q_a = (k_n^2 +k_p^2 \mp 2 k_p k_n
\cos \theta)^{1/2}/2$ and $\cos \theta =\vec k_n \cdot \vec k_p / k_nk_p$.
The corresponding difference 
of matrix elements for the three-body interaction is independent of $q$ and
is given by
\be
\Delta v^{3b} = t_3 \rho^\alpha.
\ee
These differences for nucleons on the 
Fermi surface
are plotted in Fig. 2 as a function of $\cos\theta$, taking the Fermi momentum
\begin{figure}
\includegraphics [width = 8cm]{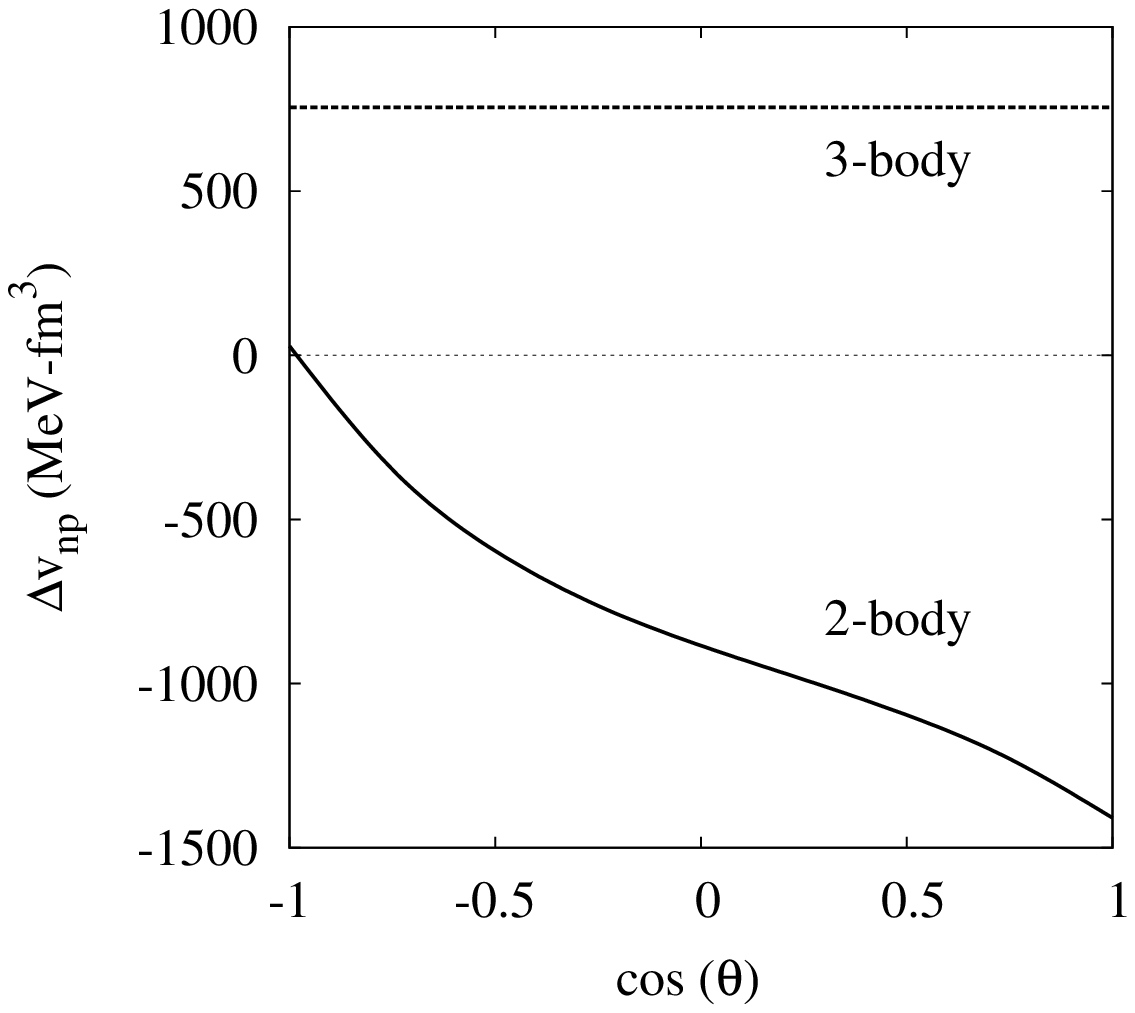}%
\includegraphics [trim = 1 0 0 0,width = 8cm]{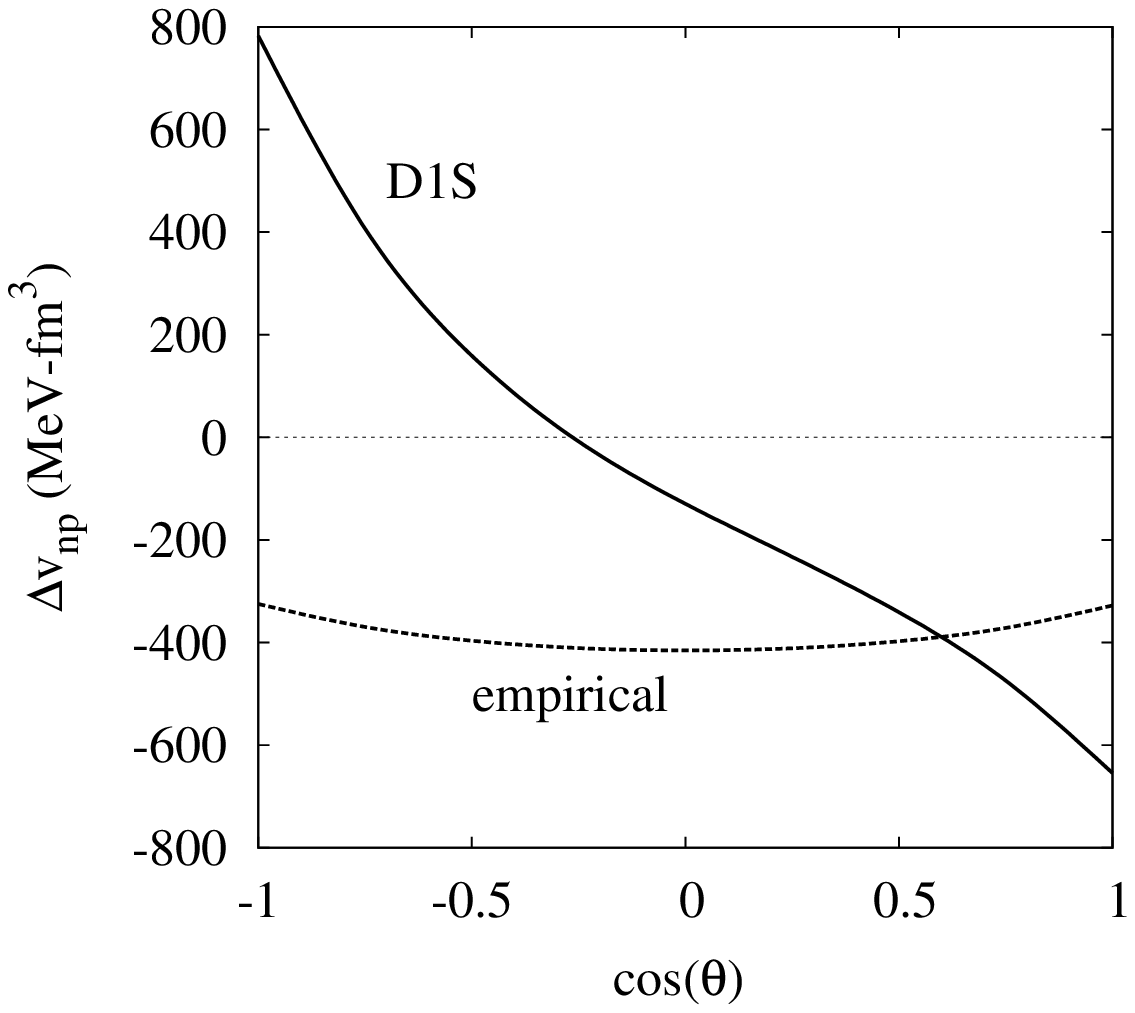}%
\caption{\label{spin-dep} Matrix elements of the effective neutron-proton
interaction from the D1S Gogny energy functional at nuclear matter density.
In the left-hand panel, the individual contributions of the two- and
three-body terms from Eq. (3) and (4) are shown.  In the right-hand
panel the total for the D1S is shown in comparison to the empirical
$\Delta v_{np}$ discussed in Refs. \cite{bo76,co97}.
}
\end{figure}
as $k_f = 1.35 $ fm$^{-1}$.
In the left-hand panel one see that the two components have opposite sign.
They are added together in the plot on the right.  Here one can
see that both signs are present, depending on the angle $\theta$.  In
that plot we also show an empirical neutron-proton interaction,
constructed to fit the data on the GM multiplets.  The interactions
are in rough agreement when the momenta in the parallel-spin state are also
parallel, but strongly disagree when the momenta are anti-parallel in that state.

We conclude with some remarks the construction of a better three-body
interaction.  It seems clear that the contact nature of the Gogny (and Skyrme)
interactions is at the root of the problem of reproducing the empirical
spin-dependence of the neutron-proton interaction.  There have been 
proposals in the literature to generalize the three-body interaction
by including derivative terms\cite{ra11,sa13} as in the Skyrme
two-body interaction.  Unfortunately the expansion in powers of the
derivatives gives rise to many terms and it is difficult from a purely
empirical point of view to determine the coefficients.  
The interaction arises both from the subnucleon degrees of freedom
that are missing from theory as well as from the correlations that are
missing from the mean-field treatment of the nucleon degrees of freedom.
The latter, called the induced three-body interaction, has a long
range \cite{gezerlisA} and a nonlocality \cite{gezerlisB} that is impossible
to take account of in a contact interaction.  A finite range has
to be introduced in some fashion, but the 
computational cost is very high.
Besides the Skyrme approach using derivatives of contact interactions, 
it may be possible to reduce the computational cost of finite-range
three-body interactions using hypercontraction\cite{pa13} or 
separable parameterizations.  We feel it would be very worthwhile to develop
codes that could apply them
 
\section{Acknowledgments} 
We thank J. Dobaczewski for comments.  The work of GFB was supported 
by the U.S. Department of Energy under Grant 
DE-FG02-00ER41132, and by the National Science 
Foundation under Grant PHY-0835543. 
The work of LMR was supported by MICINN grants Nos. FPA2012-34694, FIS2012-34479
and by the Consolider-Ingenio 2010 program MULTIDARK CSD2009-00064.

\end{document}